\documentclass[prl,amsmath,twocolumn]{revtex4}
\usepackage{amsfonts}
\usepackage{graphicx,multirow,xcolor}
\usepackage{times}
\newcommand{\tr}{{\rm Tr}}\newcommand{\mV}{{\mathcal V}}\newcommand{\mU}{{\mathcal U}}

\begin{document}
\title{Quantum Fisher Information as the Convex Roof of Variance}
\author{Sixia Yu}
\affiliation{Centre for quantum technologies, National University of Singapore, 3 Science Drive 2, Singapore 117543, Singapore}
\affiliation{
Hefei National Laboratory for Physical Sciences at Microscale and Department of Modern Physics,  University of Science and Technology of China, Hefei, Anhui 230026, China}

\begin{abstract}
Quantum Fisher information places the fundamental limit to the accuracy of estimating an unknown parameter. Here we shall provide the quantum Fisher information an operational meaning: a mixed state can be so prepared that a given observable has the minimal averaged variance, which equals exactly to the quantum Fisher information for estimating an unknown parameter generated by the unitary dynamics with the given observable as Hamiltonian. 
In particular we shall prove that the quantum Fisher information is the convex roof of the variance, as conjectured by T\'oth and Petz based on  numerical and analytical evidences, by  constructing explicitly a pure-state ensemble of the given mixed state in which the averaged variance of a given observable equals to the quantum Fisher information. 
\end{abstract}
\maketitle

Quantum Fisher information \cite{qfi} places the fundamental limit to the accuracy of estimating an unknown parameter, playing a paramount role in quantum metrology  \cite{qm1}. In the most fundamental parameter estimation task in which the parameter is generated by some unitary dynamics $U=\exp(-iH\theta)$, it holds the quantum Cramer-Rao bound $\Delta\theta F_\varrho(H)\ge 1$, in which $\Delta\theta$ characterizes the estimating accuracy by any possible measurement made on the quantum state $U\varrho U^\dagger$ and the quantum Fisher information is given by
\begin{equation}\label{def}
F_\varrho(H)=2\sum_{a,b}\frac{(\lambda_a-\lambda_b)^2}{\lambda_a+\lambda_b} |H_{ab}|^2:=4I_\varrho(H).
\end{equation}
Here we have denote $H_{ab}=\langle\psi_a|H|\psi_b\rangle$ with $\{\lambda_a,|\psi_a\rangle\}$ being the eigenvalues and corresponding eigenstates of the given mixed state $\varrho$. One fourth of the quantum Fisher information $I_\varrho(H)$ is in fact a metric adjusted skew information \cite{hans}, generalizing the skew information introduced initially by Wigner and Yanase \cite{wy}. Here we shall refer both quantities $F_\sigma(H)$ and $I_\varrho(H)$ to as the quantum Fisher information.

The quantum Fisher information $F_\varrho(H)$ is a convex function of the state $\varrho$ and it holds $F_\varrho(H)\le4\sigma_\varrho(H)$  with equality holding for pure states, where the variance is given by
\begin{equation}
\sigma_\varrho(H)=\tr\varrho H^2-(\tr\varrho H)^2.
\end{equation}
Recently T\'oth and Petz \cite{tp} found that the variance is its own concave roof, meaning that among all possible pure-state ensembles $\{p_k,|\phi_k\rangle\}$ of a given mixed state $\varrho$ there exists a special ensemble in which the averaged variance
\begin{equation}\label{av}
\sigma_\varrho^\phi(H)=\sum_kp_k\sigma_{\phi_k}(H)=\tr\varrho H^2-\sum_{k}p_k\langle\phi_k|H|\phi_k\rangle^2
\end{equation}
reaches its maximal value $\sigma_\varrho(H)$. Furthermore they conjectured that, based on some analytic results and strong numerical evidences, the quantum Fisher information $I_\varrho(H)$ is the convex roof of variance, i.e., the minimal averaged variance. In this note we shall establish this fundamental connection rigorously by constructing explicitly an ensemble in which  the averaged variance attains the quantum Fisher information, i.e.,
\begin{equation}
\min_{\phi=\{p_k,|\phi_k\rangle\}}\sigma_\varrho^\phi(H)=\frac 14F_\varrho(H).
\end{equation}

To begin with, for an arbitrary observable $H$ and a given mixed state $\varrho$ whose non zero eigenvalues are denoted by $\lambda_a$  and corresponding eigenstates by $|\psi_a\rangle$ with $a\in R$, we introduce  two related observables
\begin{eqnarray}
Z_H&=&\sum_{a,b\in R}\sqrt{\frac{2\lambda_a\lambda_b}{\lambda_a+\lambda_b}} H_{ab}|\psi_a\rangle\langle\psi_b|,\\
Y_H&=&\sum_{a,b\in R}\frac{2\sqrt{\lambda_a\lambda_b}}{\lambda_a+\lambda_b}H_{ab}|\psi_a\rangle\langle\psi_b|.
\end{eqnarray}
With the help of the first observable $Z_H$ the quantum Fisher information can be rewritten as $I_\varrho(H)=\tr\varrho H^2-\tr Z_H^2$. Let $\alpha_k$ and $|y_k\rangle$  with $k\in R$ be the eigenvalues and corresponding eigenstates for the second observable $Y_H$, which is an $r\times r$ Hermitian matrix, with $r=|R|$ being the rank of $\varrho$, in the subspace spanned by $\{|\psi_a\rangle\}_{a\in R}$. The $r\times r$ unitary matrix $U$ with matrix elements $U_{ka}=\langle \psi_a|y_k\rangle$ diagonalizes $Y_H$ as
\begin{equation}\label{U}
\sum_{k\in R}U_{ka}^*U_{kb}\alpha_k=\langle\psi_b|Y_H|\psi_a\rangle\quad (\forall a,b \in R),
\end{equation}
since $Y_H=\sum_k\alpha_k|y_k\rangle\langle y_k|$.
With the help of the unitary matrix $U$ defined above we introduce the following ensemble $\mU=\{u_k,|U_k\rangle\}_{k\in R}$ of pure states\begin{equation}\label{u0}
|U_k\rangle=\frac{1}{\sqrt{u_k}}\sum_{a\in R}U_{ka}\sqrt{\lambda_a}|\psi_a\rangle,\quad u_k=\sum_{a\in R}|U_{ka}|^2\lambda_a.
\end{equation}
Obviously we have $\varrho=\sum_{k\in R}u_k|U_k\rangle\langle U_k|$ and for each pure state $|U_k\rangle$ we can calculate the expectation value of $H$ as
\begin{equation}
\langle U_k|H|U_k\rangle=\sum_{a,b\in R}\frac{U^*_{ka}U_{kb}}{u_k}\sqrt{\lambda_a\lambda_b}H_{ab}=\frac{\tr Z_H\Gamma_k}{{u_k}}
\end{equation}
where 
\begin{equation}\label{g}
\Gamma_k=\sum_{a,b\in R}U^*_{ka}U_{kb}\sqrt{\frac{\lambda_a+\lambda_b}2}|\psi_b\rangle\langle\psi_a|
\end{equation}
are orthogonal to each other in the sense that
\begin{eqnarray}
\tr\Gamma_k\Gamma_j&=&\sum_{a,b\in R}U^*_{ka}U_{kb}U^*_{jb}U_{ja}\frac{\lambda_a+\lambda_b}2\cr
&=&\frac1{2}\sum_{a\in R}U^*_{ka}U_{ja}\lambda_a\sum_{b\in R}U_{kb}U^*_{jb}\cr
&&+\frac1{2}\sum_{a\in R}U^*_{ka}U_{ja}\sum_{b\in R}U_{kb}U^*_{jb}\lambda_b\cr
&=&\sum_{a\in R}U^*_{ka}U_{ka}\lambda_a\delta_{kj}=u_k\delta_{kj}
\end{eqnarray}
for all $k,l\in R$. Moreover from Eq.(\ref{U}) it follows the expansion $Z_H=\sum_{k\in R} \alpha_k \Gamma_k$ and thus $\tr Z_H\Gamma_k=u_k\alpha_k$ so that
\begin{equation}\label{u1}
\sum_{k\in R}{u_k}\langle U_k|H|U_k\rangle^2=\sum_{k\in R}u_k\alpha_k^2=\tr Z_H^2
\end{equation}
from which it follows that the averaged variance as defined in Eq.(\ref{av}) in the pure-state ensemble $\mU$ equals exactly to the quantum Fisher information, i.e.,
$\sigma_\varrho^\mU(H)=I_\varrho(H)$.

Any convex function of $\varrho$ that reduces to the standard variance in the case of pure states, e.g., various metric adjusted skew informations \cite{hans} as well as generalized Fisher information \cite{petz} proper scaled, is a lower bound of the averaged variance. From above discussions we know that the quantum Fisher information $I_\varrho(H)$ equals to the averaged variance in the minimal ensemble $\mU$. And thus all the other convex functions of $\varrho$ should be no larger than the quantum Fisher information, i.e., the quantum Fisher information is the maximal convex function of $\varrho$ that is the lower bound of all the possible averaged variances. Simply put,  the convex roof of variance equals to the quantum Fisher information $I_\varrho(H)$.
 
In certain sense the minimal ensemble $\mU$ for a given observable $H$ and mixed state $\varrho$ is unique as long as the unitary matrix that diagonalizes $Y_H$ is unique, or the observable $Y_H$ is non degenerated. This is  because for any unitary $U$  pure states Eq.(\ref{u0}) define an ensemble for $\varrho$ and the observables $\Gamma_k$ defined in Eq.(\ref{g}) are still orthogonal to each other. However for Eq.(\ref{u1}) to hold it is necessary for $Z_H$ to have an expansion using only orthogonal observables $\Gamma_k$, regarded as a part of a complete set of orthogonal operator basis. This determines $U$ to be the unitary matrix that diagonalizes $Y_H$. As such we obtain an operational interpretation of the quantum Fisher information and observable $H$ in a given state $\varrho$: it is the minimal averaged variance for the given observable $H$ among all possible preparations of the given mixed state $\varrho$.

For completeness we also construct an alternative pure-state ensemble $\mV$ in which the averaged variance is maximized, i.e., equals to the standard variance. 
We note that for every traceless observable $X$ there is a basis under which all diagonal elements  vanish. By an argument of continuity there exists a state $|\phi\rangle$ such that $\langle \phi|X|\phi\rangle=0$. If it is not true the observable $X$ is either positive or negative semidefinite and thus has a zero trace if and only if it equals to zero. By taking $|\phi\rangle$ as a basis and the observable $X$ in the orthogonal complement subspace is also Hermitian with zero trace and so on.

Obviously the observable $X_H=\sqrt\varrho H\sqrt \varrho-\varrho\tr\varrho H$ is traceless and there should be a basis $\{V|\psi_a\rangle\}_{a\in R}$ in which  all the diagonal elements of $X_H$ equal to zero, where $V$ is some unitary transformation, acting trivially on the zero subspace of $\varrho$. Thus  for each $k\in R$ we have $\langle\psi_k|V^\dagger X_HV|\psi_k\rangle=0$, meaning that, by denoting $V_{kb}=\langle\psi_b|V|\psi_k\rangle$, we have
\begin{equation}\label{tl}
\sum_{a,b\in R}V^*_{ka}V_{kb}\sqrt{\lambda_a\lambda_b}\Big(H_{ab}-\delta_{ab}\tr\varrho H\Big)=0.
\end{equation}
Consider now the pure-state ensemble $\mV=\{|V_k\rangle\}$ for the given state $\varrho$ where
$|V_k\rangle=\sum_{a\in R}V_{ka}\sqrt{\lambda_a}|\psi_a\rangle$ with normalization $\langle V_k|V_k\rangle=v_k$  for $k\in R$. In each pure state $|V_k\rangle$ the expectation value of $H$ can be readily calculated with the help of Eq.(\ref{tl}) to be $\langle V_k|H|V_k\rangle=v_k\tr\varrho H$, leading to 
\begin{equation}
\sum_{k\in R}\frac1{v_k}\langle V_k|H|V_k\rangle^2=\sum_{k\in R}v_k(\tr\varrho H)^2=(\tr\varrho H)^2
\end{equation}
from which it follows $\sigma_\varrho^\mV(H)=\sigma_\varrho(H)$. 

As a final remark, in a general parameter estimation task where the state is given by $\varrho=\sum_k\lambda_k|\psi_k\rangle\langle\psi_k|$ with eigenvalues $\lambda_k$ and eigenstates $|\psi_k\rangle$ depending on the unknown parameter $\theta$, the quantum Fisher information is given by $F_\varrho(\theta)=\tr\varrho L_\theta^2$ with $L_\theta$ being the symmetric logarithmic derivative satisfying $2\dot\varrho=L_\theta\varrho+\varrho L_\theta$. The quantum Fisher information has a suggestive decomposition $F_\varrho(\theta)=F_\varrho^c(\theta)+F_\varrho(H_\theta)$ \cite{paris} into  a classical Fisher information $F^c_\varrho(\theta)=\sum_k\dot\lambda_k^2/\lambda_k$ and a quantum part $F_\varrho(H_\theta)$ as defined in Eq.(\ref{def}) for the following Hamiltonian, which may be $\theta$-dependent,
\begin{equation}
H_\theta=i\sum_k|\dot\psi_k\rangle\langle\psi_k|.
\end{equation}
Our results shows that the quantum Fisher information in general, apart from a classical Fisher information arising from the $\theta$-dependence of the eigenvalues, equals to 4 times the convex roof of the variance of the Hamiltonian $H_\theta$ defined above, arising from the $\theta$-dependence of the eigenstates.

In sum, we have constructed a special ensemble for a given mixed state in which the averaged variance of a given observable attains its minimal value, which equals exactly to the quantum Fisher information, proving that the the convex roof of variance equals to the quantum Fisher information. Also we provide an alternative ensemble in which the averaged variance attains its maximal value, showing that the variance is its own concave roof as proved already in \cite{tp}.  

This work is supported by National Research Foundation
and Ministry of Education, Singapore (Grant No.
WBS: R-710-000-008-271) and NSF of China (Grant No.
11075227).

\newpage
\end{document}